\documentclass[a4paper,12pt]{article}
\usepackage{dcolumn}
\usepackage{graphicx}
\usepackage{graphics}
\usepackage{epic}
\usepackage{eepic}
\usepackage{latexsym}
\usepackage{pictex}
\usepackage{color}
\usepackage{threeparttable}
\usepackage{ifthen}
\usepackage{cite}
\begin{document}


\title{{\bf On the role of the nonlocal Hartree-Fock exchange in {\em ab initio} quantum transport: \\
H$_2$ in Pt nanocontacts revisited }}

\author{Y. Garc\'{\i}a\thanks{E-mail: Yamila.Garcia@uv.es} \\
Instituto de Ciencia de Materiales, \\
Universidad de Valencia, E-46071 Valencia, Spain \\ 
\\
and \\
\\
J.C. Sancho-Garc\'{\i}a \\
Departamento de Qu\'{\i}mica-F\'{\i}sica, \\
Universidad de Alicante, E-03080 Alicante, Spain.
}

\date{\today}

\maketitle

\vspace{-14cm}
\begin{flushright}
\end{flushright}
\vspace{13cm}

\begin{abstract}
\setlength{\baselineskip}{0.333333333in}
We propose a practical way to overcome the ubiquitous problem of the overestimation of the 
zero-bias and zero-temperature conductance, which is associated to the use of local approximations 
to the exchange-correlation functional in Density-Functional Theory  when applied to quantum transport.
This is done through partial substitution of the local exchange term in the functional by the 
nonlocal Hartree-Fock  exchange. As a non-trivial example of this effect we revisit the smallest 
molecular bridge studied so far: a H$_2$ molecule placed in between Pt nanocontacts.
When applied to this system the value of the conductance diminishes as compared to the 
local-exchange-only value, which is in close agreement with results predicted from Time-Dependent 
Current-Density-Functional Theory. Our results issue a warning message on recent claims of perfect 
transparency of a H$_2$ molecule in Pt nanocontacts.
\end{abstract}


\setlength{\baselineskip}{0.333333in}

It is widely admitted that a first-principles methodology based on Density-Functional Theory (DFT)
is the only compromise between accuracy and computational resources that can help pave the way 
towards molecular-engineered nanoscale devices\cite{Kohn:essay}. The DFT manageability is based, for 
the most part, in the use of  the Kohn-Sham scheme (DFT-KS)\cite{Hohenberg:pr:64,Kohn:pr:65}.
In addition, the central track doing a practical DFT-KS scheme lies on the definition of an approach 
to the true exchange-correlation (xc) functional\cite{Nesbet}. Initially, the manageability of 
DFT-KS comes upon the introduction of the Local-Density Approximation (LDA) to the xc 
functional\cite{Kohn:pr:65}. (Henceforth, extensions beyond LDA within local exchange models, 
like generalized gradient approximations (GGA)\cite{perdew_gga}, and Meta-GGA\cite{tpss}, are 
referred as semi-local methods and labeled as LDA methods for the sake of simplicity).

This {\em ab-initio} based method  combined with Landauer's formalism is nowadays routinely 
applied to compute the conductance, $G$, of nanoscaled systems, i.e metallic nanocontacts and 
single-molecule junctions\cite{Agrait:pr:03,Palacios:ctcc:05,jc.add}. However, the use of 
DFT-KS electronic structure  within LDA approximations, would present two  problems 
in this regard: (i) an obvious one related to the use of LDA and related approximations to 
approach the true xc-potential, $v_{xc}$\cite{Nesbet,jc.add}; and (ii) a much more subtle one which
could be associated with the absence of dynamical corrections in the  DFT-KS 
schemes\cite{Sai:prl:05,vignale.kohn}. 

Regarding the first point, the reliability of LDA results relies naturally on the existence of
fairly homogeneous, well-behaved electronic densities. While this is the case for metallic 
nanocontacts, molecular bridges, definitely, do not satisfy this premise\cite{Sai:prl:05}.
For weakly-coupled molecules conducting in the Coulomb blockade regime, where charge localization 
is strong, failures of LDA are traced back to the self-interaction and derivative discontinuity 
problems, which are inherent to local or semi-local 
approximations\cite{palacios:prb:05,evers,sanvito,jc.add}.
Interestingly, this problem is not present in more sophisticated exact exchange (EXX) nonlocal
approaches, as recently emphasized by others\cite{jc.add2}.

Concerning the second point, the use of Time-Dependent Density-Functional Theory (TDDFT)\cite{jc.add3}
has been shown to give the exact total current if the exact $v_{xc}$ is known\cite{Sai:prl:05}, and
it reduces in the zero-frequency limit to the standard static treatment only when the adiabatic 
approximation is invoked\cite{Sai:prl:05}. Furthermore, as initially suggested by Vignale and 
Kohn\cite{vignale.kohn}, the apparent impossibility to approach the exact $v_{xc}$ through 
successive local approximations could be circumvented by a new form of functional where the local 
current density plays a central role or, equivalently, a nonlocal density-based theory.
Recent work\cite{Sai:prl:05} based on Time-Dependent Current-Density-Functional Theory 
(TDCDFT) has shown that dynamical effects manifest themselves in the appearance of a dynamical 
potential, $v_{\rm xc}^{\rm dyn}$, that always opposes the electrostatic potential and lowers the 
conductance obtained within LDA methods, even in linear response. The origin of this potential can 
be traced back to the nonlocal response to an electric field which is captured by 
TDCDFT\cite{vignale.kohn,Sai:prl:05}. It is difficult to assess to what extent the resulting
dynamical potential accounts for the lack of nonlocality of the starting LDA approximation or 
if it represents a true intrinsic dynamical effect which is present even for the exact DFT-KS 
potential\cite{jc.add} (probably both).

The role played by the dynamical term in increasing the  resistance is compatible with another 
seemingly unrelated fact: The polarizability of molecular systems (e.g., hydrogen\cite{St:Kr:Perdew} or 
polymeric\cite{tdcdft} chains) is severely overestimated by LDA, while Hartree-Fock (HF), 
TDCDFT, and exact many-body calculations --all of them nonlocal schemes-- yield similar values. 
We note here that a recent work has shown that the conductance can be expressed in 
terms of the polarizability\cite{godby}.

In the light of the above discussion, one might ask: {\em to what extent can nonlocal approaches 
to the $v_{xc}$ (out of the realm of standard static DFT-KS theory) mimic the missing dynamical 
corrections?} In this work we explore the influence of a nonlocal potential -such as HF
exchange- on the conductance of a molecular bridge (see Figure 1). To this end we choose
to study electron transport in a H$_2$ molecule bridging Pt 
nanocontacts\cite{Smit:nature:02,Garcia:prb:04,Thygesen:prl:05}
(see inset in Figure 2), using nonlocal functionals customized for this problem.
We show that the HF-like exchange has a strong influence on the conductance of this bridge,
rapidly decreasing as the percentage of HF-like increases in the hybrid functional.
By adjusting this percentage, with the aid of accurate quantum chemistry calculations in clusters, 
we conclude that previously reported calculations based on LDA might have systematically overestimated 
the value of the conductance in this system. Finally, for this benchmark system, we compare the 
numerical results derived from our method, the nonlocal functionals for an effective DFT-KS formalism,
with those obtained within TDCDFT\cite{Sai:prl:05}. The conductance values are quite similar, which 
further supports the conclusions reached here. The average value of the conductance for such system 
turns out to be $\approx 0.2 \times 2e^2/h= 0.2 \times G_0$, in stark contrast to LDA calculations 
yielding $\approx 1.0 \times G_0$ and in support of  a previous work by some of us\cite{Garcia:prb:04}.

Our theoretical approach have been thus developed in accordance with the scheme that follows.
The central track doing an effective DFT-KS scheme lies on the definition of the  
$v_{xc}$\cite{Kohn:essay,martin.book,Nesbet}. Rooted on the adiabatic connection theorem, which 
formally justified the design of nonlocal functionals (usually referred as hybrid 
functionals\cite{adamo}), and based on the experience gained after original Becke's 
proposal\cite{Becke:jcp:93}, we adopt the following expression for an approach to the  
$v_{xc}$:
\begin{equation}
\label{mixing}
v_{xc} = \alpha  v_{x}^{HF} + \left( 1 - \alpha \right) v_{x}^{local} + v_{c}^{local},
\end{equation}
where $v_{x}^{HF}$ is the  HF-like potential contribution, which accounts for nonlocal effects, the 
$v_{x,c}^{local}$ contributions are given by any of the available exchange (x) or correlation (c) 
local functionals and, finally, $\alpha$ is a fitting parameter. With this parameterized potential, 
we will solve the mean-field equations derived from DFT-KS formalism for a finite system, the cluster.
This finite cluster is selected in such a way that the boundary conditions determined by the problem
are fulfilled, as discussed in the next paragraphs. Finally, the electronic states derived from the 
DFT-KS are deployed in conjunction with Landauer formalism for zero temperature quantum transport 
characterization\cite{Datta:book:95}. Concerning the software employed for the numerical calculations, 
we use the code ALACANT (ALicante Ab initio Computation Applied to NanoTransport)\cite{Louis:prb:03} 
which is interfaced to GAUSSIAN03\cite{Gaussian:03}.

We first compute the LDA conductance ($\alpha=0$) of the system shown in the inset of 
Figure 2 where the  H$_2$ molecule lies along the Pt nanocontacts axis\cite{basis}. In agreement with 
previous works\cite{Smit:nature:02,Cuevas,Thygesen:prl:05}, we obtain a single fully conducting 
channel at the Fermi level which  yields $G \approx  1.0 \times G_0= 2 e^2/h$. This result supports the
implementation for the computation of the current\cite{Palacios:prb:02}.

To probe in a systematic way the influence of the nonlocal  exchange on the electronic structure and 
concomitant change in the conductance, the coefficient $\alpha$ is now varied. Strictly speaking 
$\alpha$ should not be a constant, but a function of spatial coordinates, $\alpha = \alpha(\vec{r})$.
It is expected that  HF-like exchange is more needed to describe electronic density regions governed 
by just one electron where the self-interaction error is more notorious\cite{Scuseria}.
This is particularly the case for the H$_2$ molecule. The value of $\alpha$, on the other hand, 
should decrease as we move into the bulk electrodes where LDA  or GGA is expected to perform better. 
As a compromise between both situations we take $\alpha$ to be constant in the nanocontacts region 
close to the H$_2$ molecule where the coordination is low, i.e., the bridge  Pt-H$_2$-Pt (see inset 
in Figure 3). The rest of the system is described by the self-energy corresponding to 
a parameterized Bethe lattice. Figure 1 shows the effect of the HF-like exchange on the 
conductance. For values close to 50\% the conductance has almost dropped to zero.


The drop in conductance as the amount of HF-like exchange increases at the bridge can also be 
understood from a purely microscopic point of view. This construction is based on the approach to the 
exact non-correlated density-based scheme for determine the electronic structure\cite{martin.book},
and its combination with the accepted Landauer formalism for determine the electronic 
conductance\cite{Agrait:pr:03}. Therefore, under these two conditions, the electronic structure 
limited to the  HOMO-LUMO gap, $\Delta_{HOMO-LUMO}$, will play a key role in the final outcome of 
the electronic conductance\cite{datta.ghosh}. Figure 3 shows the effect of the HF-like 
contribution on the HOMO-LUMO gap ($\Delta_{HOMO-LUMO}$) for the isolated Pt-H$_2$-Pt bridge.
This effect is well known in molecules and bulk semiconductors where LDA approximations usually 
underestimate the real value of the charge gap\cite{martin.illas,gap_Si,bandgapssemiconductors}.
The one electron  Green's function ($ g(\epsilon)$) of the isolated Pt-H$_2$-Pt  bridge
has poles at the position of the molecular orbitals energies. Once the isolated system is coupled 
to the rest of the electrodes these poles shift and broaden in a way dictated by the self-energies 
$\Sigma_{\rm L}$ and $ \Sigma_{\rm R}$ representing the electrodes:
\begin{equation}
\label{green}
 g^{(\pm)}(\epsilon)=[(\epsilon\pm\,i\delta){ I}-{H} - \Sigma_{\rm L} - \Sigma_{\rm R}]^{-1}.
\end{equation}
The Fermi level is likely to lie in the gap between the HOMO and LUMO states of the Pt-H$_2$-Pt 
cluster since no significant charge transfer is expected between Pt atoms. As can be seen in 
Figure 3, the LDA gap is almost zero for this bridge. From the opening of the gap as $\alpha$ 
increases one could easily have anticipated a strong decrease in conductance since this depends on the
Green's functions  through the well-known expression\cite{Datta:book:95}:
\begin{equation}
\label{G.def}
G=\frac{2e^2}{h} Tr[\hspace{2 mm}{\Gamma_L}(\epsilon)\hspace{2 mm}{ g^+}(\epsilon)\hspace{2 mm}{\Gamma_R}(\epsilon)\hspace{2 mm}{ g^-}(\epsilon)].
\end{equation}

Finally, to estimate $\alpha$ we follow a well-established methodology in semiconductors and 
insulators.
The value is chosen so that $\Delta_{HOMO-LUMO}$ matches the charge gap, $GAP$. Accordingly, it is 
defined as $GAP = 2 E_0(N) - E_0(N+1) - E_0 (N-1)$, where $N$ represents the number of electrons, and 
$E_0$ is the total energy for the selected ground states
in  a Kohn-Sham system of {\em independent  electrons}.
Matching between $\Delta_{HOMO-LUMO}$ and 
$GAP$ will be enough to guarantee accuracy for mean-field Green's 
functions\cite{bickelhaupt.1,spectral}, and therefore for the conductance value, expressions 
\ref{green} and \ref{G.def}. 
The choice of the finite region where to perform the search of $\alpha$ 
is constrained  by three conditions: (i) it must contain the zone which is restricting the current 
flow, in our case the H$_2$ molecule; (ii) the HOMO and LUMO must not be localized on the electrodes;
and (iii) screening effects from the contacts should be -at least- partially included.
These three conditions reduce the range of possibilities to the cluster shown in Figure 3,
and limit the selection of system (previously referred as the  extended 
molecule\cite{jc.add,Nitzan:science:03}). One way to estimate to what extent screening is included in 
such small cluster is to look for a quantity such as the  internal mode frequency of the H$_2$ 
molecule, $w_{H_2}$, and check how much it changes with increasing the size of the cluster. We find 
that larger clusters do not appreciably change that quantity, which is given by
$w_{H_2} \approx w_{H_2}^0/2$ where $w_{H_2}^0$ is the internal mode of an isolated H$_2$. This change 
in the frequency with respect to the free molecule suggests an intermediate regime of coupling.

To find out ground state energies and therefore the  right $GAP$, we begin by considering 
the mean-field  HF method. To approach the exact many-body solution to the $GAP$,
we invoke hierarchical correlated {\em ab initio} methods such as M\"oller-Plesset perturbation 
(MP) or Coupled-Cluster (CC) theories. The corresponding results are presented in Figure 4.
The $GAP$ for this system, $\sim 5$ eV, critically differs from the 
$\Delta_{HOMO-LUMO}$, described from the electronic structure in pure DFT-LDA calculations 
$\sim 0$ eV, see Figure 3. Therefore, and considering the scheme we propose to correct 
this pitfall, the  $\Delta_{HOMO-LUMO}$ for the system is  extracted from Figure 4.
Comparing those results with the whole information contained in curve of Figure 3,
we estimate that the percentage of HF exchange in Eq. \ref{mixing} should be  $\sim 60$ percent, 
which gives 
 a very low  value for the conductance (see Figure 1).
It is the main result we should draw from this paper.

Note that there is an increasing concern about the influence of exact exchange for conduction
phenomena at the molecular level\cite{add3,add4}. Although a rule of thumb for the needed admixture 
of exact exchange is provided here, we indeed believe that it will be difficult to unambiguously give 
the optimum weight of exact exchange for an uniformly good description of all systems. Instead of 
this, we would like to remark how the results clearly show that inclusion of exact exchange is 
necessary for accurate descriptions. 
We should point that the same amount of HF exchange is obtained (with error band of 10 percent), if the
contour condition to find out its value is limited to the  matching between  the  HOMO eigenvalue and the negative of 
the ionization potential, $\epsilon_{HOMO} = -IP$\cite{add8}. 
Concerning future work and despite being technically 
challenging, it would be definitively interesting to investigate the performance on these benchmark 
systems of a new generation of functionals describing $v_{xc}$. For instance, we should mention 
orbital-based functionals\cite{add5}, self-interaction corrected functionals\cite{add6,sic.portal},
the hybrid functionals based on a screened Coulomb potential\cite{ad1,ad2}, and the last generation
of $v_{xc}$ density-functionals belonging to the M06 suite of methods\cite{ad3,ad4}.
Furthermore, more studies for closely related molecules and nanocontacts will be highly welcomed too.

If  we were able to follow unambiguously  the same procedure for larger systems the value for 
$\alpha$ would probably be reduced, although we do not expect it to go to zero. Unable to get rid of 
this uncertainty we compare our results to those obtained with TDCDFT following the work by 
Sai et al.\cite{Sai:prl:05}. A dynamical counteracting potential appears whenever there is a change 
in the macroscopic density which modifies the local-density conductance: 
\begin{equation}
G=\frac{G_{\rm local}}{1+G_{\rm local} R^{\rm dyn}}
\end{equation}
From the average density profile along the z-direction  (see Figure 5) we compute the 
corrections to the local conductance. The electronic density is fixed in the electrodes to the bulk 
value, considering for the number of free-electrons as the mean experimental atomic-point-contact 
transmission. In case of Pt-nanocontacts, this number implies a range between one- and 
two-free-electrons per Pt atom\cite{Smit:nature:02}. The rest of the parameters involved in the 
calculation of $R^{\rm dyn}$ were estimated in agreement with the method proposed by 
Sai et al.\cite{Sai:prl:05}. Our calculations brings  the conductance down to $\approx 0.2 \times 
G_0$. (Details are included in Ref. \cite{Y:thesis:07}). Nevertheless, this procedure is not free from 
ambiguities either, but it points towards similar conclusions as the ones drawn above.

In summary, we have presented a route to approach the  DFT-KS  electronic transport calculations 
based on the {\em controlled} addition of EXX contribution to $v_{xc}$. The selection of the 
fraction of HF-like needed is based on first-principles procedures. Our results agree with the ones 
obtained with the method developed by Sai et al\cite{Sai:prl:05} within TDCDFT, and supports the 
functionality of DFT methods based on nonlocal models for $v_{xc}$. Its agreement
opens up  possibilities to formally justify a connection between {\em nonlocal} DFT and TDCDFT in the 
zero-frequency limit. Furthermore, more work is needed on this respect since the  exact amount of nonlocality might 
be difficult to estimate given: (i) the  difficulties for the definition of the region where
it should operate; and (ii) the correct implementation of this behavior.
However, we can at least define an upper limit for this nonlocal HF correction,
and give arguments in favour of its non-zero value, as DFT-LDA methods considered.
Within this context, the numerical calculation  of the conductance for a H$_2$ molecule in Pt 
nanocontacts have been corrected, lowering in one order of magnitude the standard DFT-LDA calculation.
Note, however, that the statements and the conclusions that will be made are not limited to 
calculations of conductance in this benchmark system. It is widely accepted that nonlocal 
contributions to $v_{xc}$  do play a significant role to get beyond DFT-LDA results either in 
molecular systems or bulk materials. For instance, we could mention the polarizability of polymeric
chains\cite{tdcdft}, the magnetic coupling and the band structure in systems such as
 antiferromagnetic insulators\cite{illas11,illas.22} and magnetic oxides\cite{ricardito}.
Finally, our theory is that {\em the overestimation of the conductance is not intrinsic to the DFT-KS 
approach}. This wrong behavior could be greatly improved by a controlled design of an exchange-correlation potential 
which incorporates  nonlocal contributions. 

We gratefully acknowledge M. DiVentra's detailed revision of this manuscript. Y. Garc\'{i}a is 
indebted to Professor Francesc Illas for recommending the use of quantum chemistry methods,
and also acknowledges valuable discussions with   J.J. Palacios and E.Louis (Alicante, Spain)
along last years. J.C.S.G. also thanks the ``Ministerio de Educaci\'on y Ciencia'' 
(CTQ2007-66461/BQU) of Spain for a research contract under the ``Ram\'on y Cajal'' program and 
the ``Generalitat Valenciana'' for further economic  support.

\clearpage

\begin{itemize}
\setlength{\baselineskip}{0.333333in}

\item {\bf Figure 1.} 
Reduction of the conductance due to the progressive substitution of the local xc-potential by 
the nonlocal HF-like contribution. The inset shows the conductance as function of energy calculated 
in a nonlocal DFT approximation for $\alpha =0.6$.

\vspace{.5cm}

\item {\bf Figure 2.} 
Transmission as  a function of energy calculated in a local DFT approximation for the Pt-H$_2$-Pt 
bridge shown in the inset.

\vspace{.5cm}

\item {\bf Figure 3.} 
Enhancement of the HOMO-LUMO gap ($\Delta_{HOMO-LUMO}$) due to the substitution of local xc 
potential by the nonlocal HF contribution. The inset shows  the Pt-H$_2$-Pt system  bridging the 
Platinum electrodes.

\vspace{.5cm}

\item {\bf Figure 4.} 
Ground state energy calculations of the charge gap, $GAP$, for the system shown in the inset.
$GAP = 2 E_0(N) - E_0(N+1) - E_0 (N-1)$, where N represents the number of electrons, and $E_0$ is 
the total energy for the selected ground states.
Total energy calculations were done using wave-function based methods: Hartree-Fock (HF),
M\"oller-Plesset perturbation theory at the second, third and fourth order (MP2, MP3, MP4, respectively) 
and Coupled-Cluster theory with single, doubles and single, doubles, and perturbatively estimated 
triples [CCSD, CCSD(T), respectively] methods.

\vspace{.5cm}

\item {\bf Figure 5.} 
Electron density profile after planar average (dashed line) and three-dimensional average of the 
microscopic density.

\end{itemize}

\clearpage
\oddsidemargin 0.8cm

\begin{figure}
\begin{center}
\scalebox{0.55}{\includegraphics{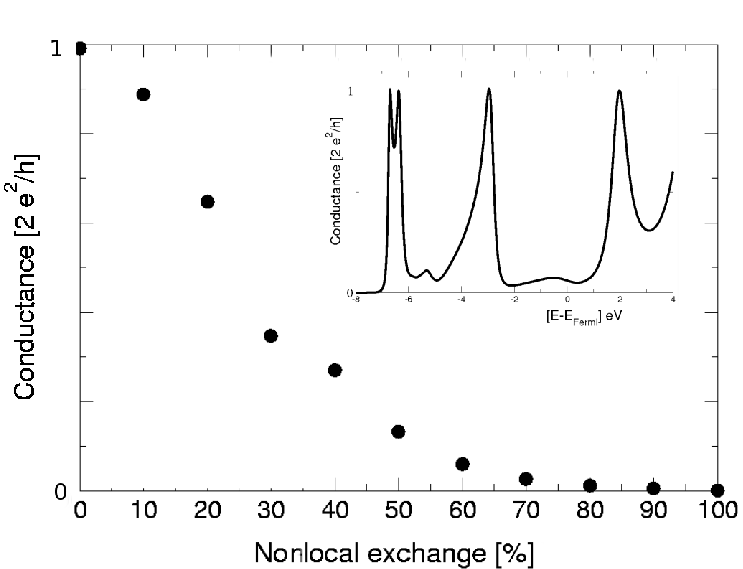}} \\
\vspace{1cm} \hspace{.5cm}
Figure 1
\label{fig:Fig1}
\end{center}
\end{figure}

\clearpage  
\oddsidemargin 0.8cm

\begin{figure}
\begin{center}
\scalebox{0.35}{\includegraphics{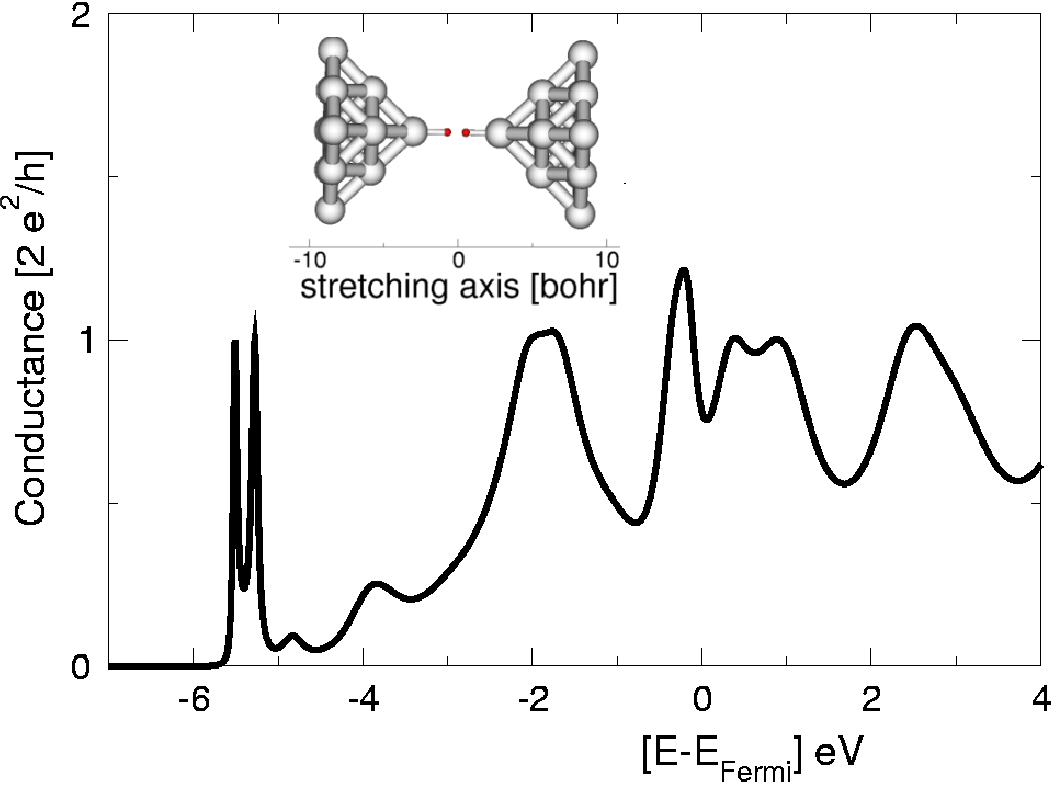}}
\vspace{1.2cm} \hspace{.5cm}
Figure 2
\label{fig:Fig2}
\end{center}
\end{figure}

\clearpage
\oddsidemargin 0.8cm

\begin{figure}
\begin{center}
\scalebox{0.31}{\includegraphics{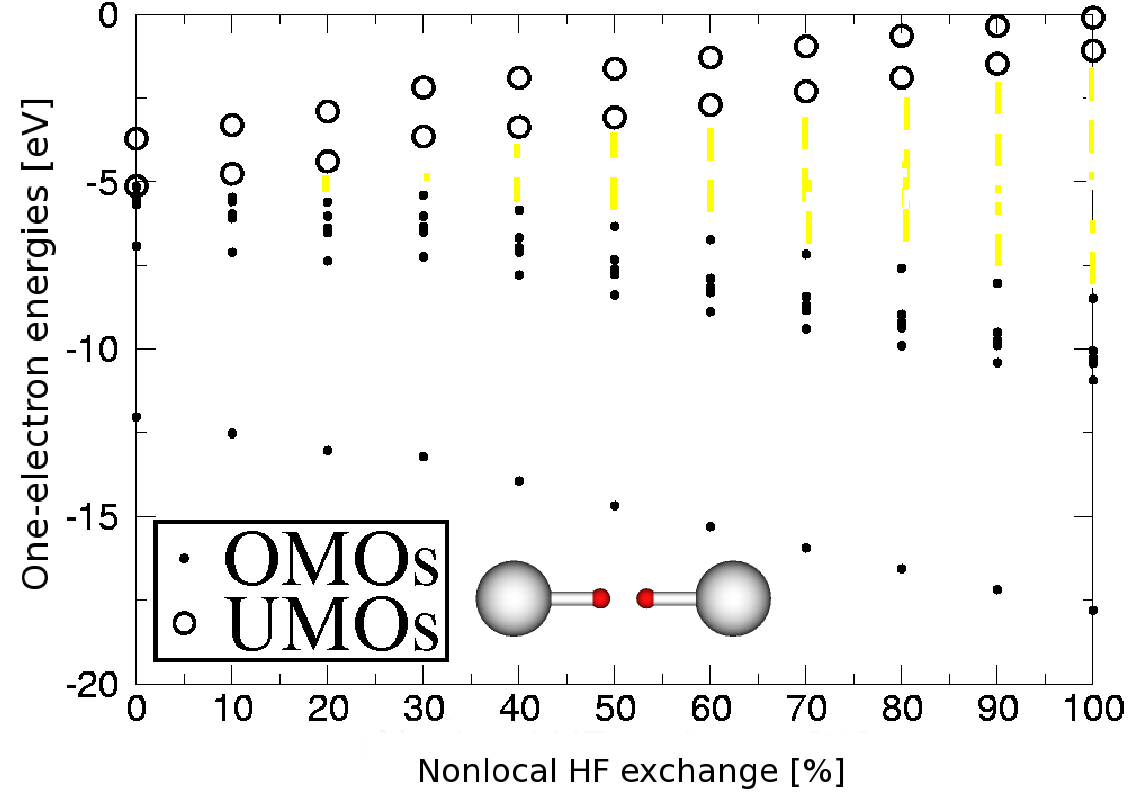}}
\vspace{1.2cm} \hspace{.5cm}
Figure 3
\label{fig:Fig3}
\end{center}
\end{figure}

\clearpage
\oddsidemargin 0.8cm

\begin{figure}
\begin{center}
\scalebox{0.40}{\includegraphics{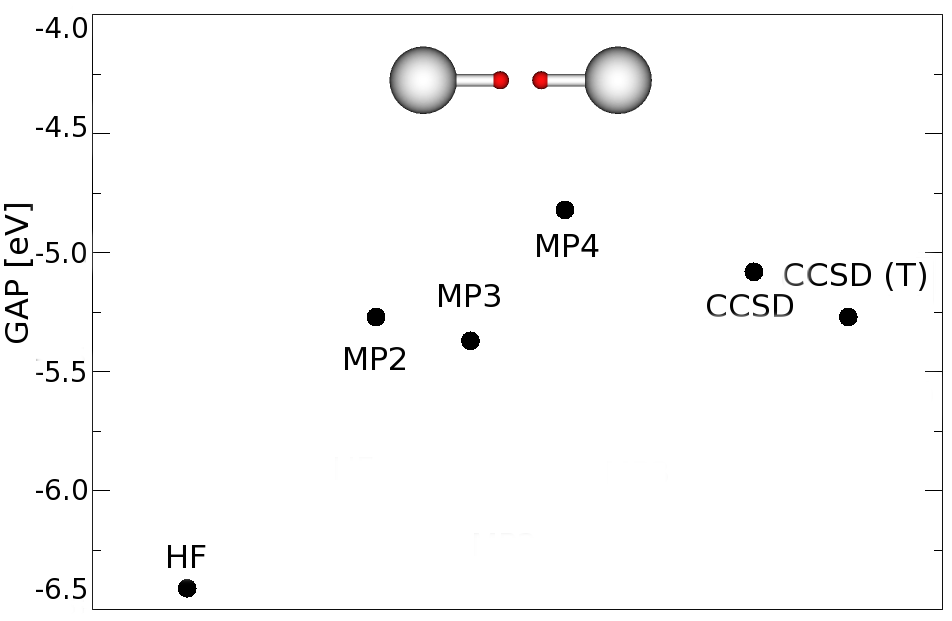}}
\vspace{1.2cm} \hspace{.5cm}
Figure 4
\label{fig:Fig4}
\end{center}
\end{figure}

\clearpage
\oddsidemargin 0.8cm

\begin{figure}
\begin{center}
\scalebox{0.5}{\includegraphics{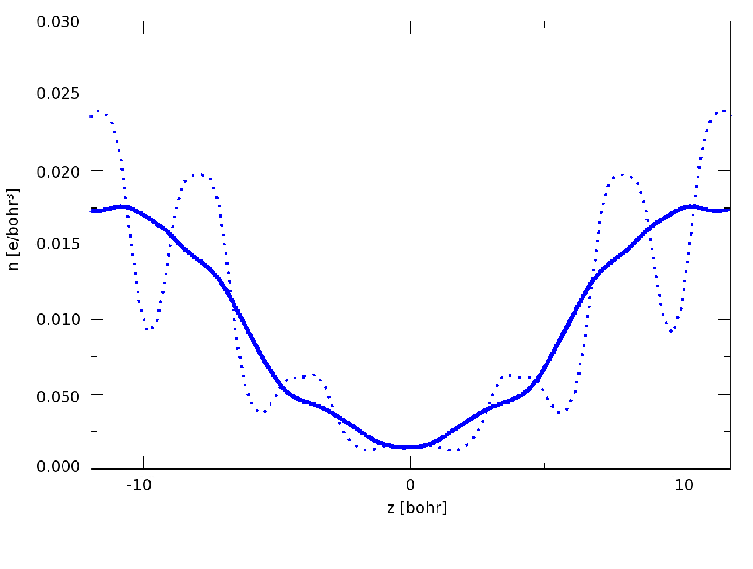}}
\vspace{1.2cm} \hspace{.5cm}
Figure 5
\label{fig:Fig5}
\end{center}
\end{figure}


\begin{thebibliography}{53}
\bibitem[29]{basis}
\bibinfo{note}{The molecular structure has been optimized using a DFT-LDA
  approach with localized atomic-orbital basis sets for Pt (lan2dz) and H
  (cc-pVTZ). Subsequently, an enhanced molecular bridge Pt-H$_2$-Pt ($^{1}\Sigma_g$)
  can be defined with a Pt-H (H-H) distance of 3.2 (1.7) bohr. Dipersions from
  these values due to different selections of exchange-correlation functionals
  and basis sets are expected not to be relevant for the conductance
  calculations further performed. These results agree fairly well with previous
  works\cite{Smit:nature:02,Garcia:prb:04,Thygesen:prl:05}}.
\end{thebibliography}
\end{document}